\documentstyle[12pt,epsf]{article}

\newlength{\extraspace}
\newlength{\extraspaces}
\catcode`\@=11

\def\numberbysection{\@addtoreset{equation}{section}
\def\theequation{\arabic{section}.\arabic{equation}}}

\newcommand{\be}{\begin{equation}
\addtolength{\abovedisplayskip}{\extraspaces}
\addtolength{\belowdisplayskip}{\extraspaces}
\addtolength{\abovedisplayshortskip}{\extraspace}
\addtolength{\belowdisplayshortskip}{\extraspace}}
\newcommand{\ee}{\end{equation}}
\newcommand{\ba}{\begin{eqnarray}
\addtolength{\abovedisplayskip}{\extraspaces}
\addtolength{\belowdisplayskip}{\extraspaces}
\addtolength{\abovedisplayshortskip}{\extraspace}
\addtolength{\belowdisplayshortskip}{\extraspace}}
\newcommand{\ea}{\end{eqnarray}}
\begin{document}
\addtolength{\baselineskip}{.7mm}
\thispagestyle{empty}\begin{center}
\begin{flushright}
TIT/HEP--299 \\
{\tt hep-th/9509143} \\
September, 1995
\end{flushright}
\vspace{3mm}
\begin{center}
{\Large
{\bf  Vacuum Energies and Effective Potential \\
in Light-Cone Field Theories }
} \\[18mm]
{\sc Shin-ichi Kojima},\footnote{
\tt e-mail: kotori@th.phys.titech.ac.jp} \hspace{2.5mm}
{\sc Norisuke Sakai}\footnote{
\tt e-mail: nsakai@th.phys.titech.ac.jp} \hspace{2.5mm}
and \hspace{2.5mm}
{\sc Tadakatsu Sakai}\footnote{
\tt e-mail: tsakai@th.phys.titech.ac.jp} \\[3mm]
{\it Department of Physics, Tokyo Institute of Technology \\[2mm]
Oh-okayama, Meguro, Tokyo 152, Japan} \\[4mm]
\end{center}
\vspace{18mm}
{\bf Abstract}\\[5mm]
{\parbox{13cm}{\hspace{5mm}
Vacuum energies are computed in light-cone field theories
to obtain effective potentials which determine vacuum condensate.
Quantization surfaces interpolating between the light-like
surface and the usual spatial one are useful to define the vacuum energies
unambiguously.
The Gross-Neveu, $SU(N)$ Thirring, and $O(N)$ vector models
are worked out in the large $N$ limit.
The vacuum energies are found to be independent of the interpolating
angle to define the quantization surface.
Renormalization of effective potential is explicitly performed.
As an example of the case with nonconstant order parameter,
two-dimensional QCD is also studied.
Vacuum energies are explicitly obtained in the large $N$ limit which
give the gap equation as the stationary point.
}}
\end{center}
\vfill
\newpage
\vfill
\newpage
\setcounter{section}{0}
\setcounter{equation}{0}
\setcounter{footnote}{0}
\numberbysection

\vspace{7mm}
\pagebreak[3]
\addtocounter{section}{1}
\setcounter{equation}{0}
\setcounter{subsection}{0}
\setcounter{footnote}{0}
\begin{center}
{\large {\bf \thesection. Introduction}}
\end{center}
\nopagebreak
\medskip
\nopagebreak
\hspace{3mm}

Quantization on light-cone has been proposed to offer a
nonperturbative method for field theories
\cite{Dirac}, \cite{tHooft1/N}.
It is relatively easy to identify genuine dynamical
degrees of freedom in this method.
One of the basic reason for this simplicity is kinematical:
light cone momentum $p_+={p_0+p_1 \over \sqrt{2}}$ for a particle
is always positive.
Therefore the particle-antiparticle pair condensation is forbidden
by the light-cone momentum conservation alone.
Therefore the vacuum in the light-cone limit is apparently the
trivial Fock vacuum. 
By virtue of the trivial vacuum, one can easily compute, for example,
mass spectra and wave functions \cite{PauliBrodsky}.
To derive these quantities more efficiently,
discretized light-cone method or light-cone Tamm-Dancoff method
have been proposed and have produced interesting results
\cite{HoBr}, \cite{MaSaSa}.

On the other hand, there are some drawbacks in the light-cone
field theories.
Firstly, loss of manifest covariance generally complicates
the renormalization procedure of light-cone field theories,
since counterterms are no longer restricted by the covariance
\cite{MaYa}.
More importantly, it is difficult to uncover the vacuum structure
such as the vacuum condensate or the spontaneous symmetry breaking.
The question of vacuum structure is usually analyzed in terms of
zero mode constraints \cite{BePi}.
To explain zero mode analysis for spontaneous symmetry breaking,
let us consider the scalar $\phi^4$ model in two dimensions
in light-cone coordinates $x^+={x^0+x^1 \over \sqrt2}$ and
$x^-={x^0-x^1 \over \sqrt2}$,
\begin{equation}
{\cal L}=\partial_+\phi \partial_-\phi - V(\phi),
\qquad V(\phi)={m^2 \over 2}\phi^2+{\lambda \over 4!}\phi^4.
\end{equation}
Defining the canonical momentum $\pi$ introduces a primary constraint
\begin{equation}
\pi\equiv{\partial {\cal L} \over \partial\partial_+\phi}=\partial_-\phi.
\label{primaryconst}
\end{equation}
By adding the constraint with an arbitrary coefficient $v(x)$, the
Hamiltonian is given by
\begin{equation}
{\cal H}=\pi \partial_+\phi -{\cal L}
+ v \cdot \left(\pi-\partial_-\phi\right)
= V(\phi)
+ v \cdot \left(\pi-\partial_-\phi\right).
\end{equation}
Since the nonzero modes of the primary constraint (\ref{primaryconst})
is of second class,
the time evolution of the primary constraint determines $v(x)$
\begin{equation}
\left[(\pi-\partial_-\phi)(x), {\cal H}(y)\right]
\delta(x^+ - y^+)
=-i
\left({d V \over d\phi} (x) +2\partial_-v(x)\right)\delta^{(2)}(x-y)
\approx 0.
\label{secondary}
\end{equation}
However, the zero mode part of the Eq. (\ref{secondary}) gives the
secondary constraint which is called the zero mode constraint
\begin{equation}
0=\int_{-L}^Ldx^-
{d V \over d\phi} (x)
=
m^2\int_{-L}^L dx^- \phi(x) +
{\lambda \over 3!}
\int_{-L}^L dx^- \phi^3(x),
\label{zeroconst}
\end{equation}
where we compactify the spatial direction and impose a periodic
boundary condition to define the zero mode of
$\phi$ unambiguously.
Eq. (\ref{zeroconst}) shows that the zero mode is not an independent
variable, but is given as a nonlinear expression of nonzero modes.
By laboriously analyzing the constraint (\ref{zeroconst}), one can
find a solution with nonvanishing zero modes in certain cases
which  indicates the spontaneous symmetry breaking \cite{BePi}
\begin{equation}
\int_{-L}^Ldx^- \phi(x)\not=0.
\end{equation}

One should distinguish two kinds of zero modes.
One type is the zero mode associated to the above constraint.
The other zero mode is the dynamical zero mode of gauge fields
which arises because of the nontrivial topology due to
the compactified spatial dimensions \cite{HeHo}.
The former is directly related to the question of vacuum condensate or
the spontaneous symmetry breaking, whereas the latter is often be
responsible to nonperturbative effects associated to the
gauge fields .
There has also been a number of works aiming at determining vacuum
structures with methods like Hartree type equations in the light-cone
field theories \cite{Lenz}.
It has been proposed to use regularizations to define
the light-like quantization surface as a limit of space-like
surfaces \cite{Lenz}, \cite{Horn}.
On the other hand, the most efficient method to find the vacuum
condensate in the covariant approach is usually to compute vacuum
energies and to obtain the effective potential
\cite{CoWe}--\cite{Abbott}.
More complicated models such as the
two-dimensional QCD coupled to quarks in the fundamental
representation are also studied in the light-cone gauge using the
large $N$ limit \cite{THooft2D}--\cite{BG}.
It has been observed that the chiral symmetry breaking occurs
in the large $N$ limit and the quark-antiquark condensation
has been computed  \cite{Zhitni}--\cite{Li}, \cite{Lenz}.
Higher order corrections in the $1/N$ expansion \cite{Witten}
convert this spontaneous symmetry breaking to
the Berezinski-Kosterliz-Thouless phenomenon \cite{KT} and make
the result consistent with the Coleman's theorem \cite{Cole73}.

The purpose of our paper is to compute vacuum energies explicitly
in light-cone field theories and to demonstrate that the effective
potential can be obtained to determine nontrivial vacuum condensate.
In order to define the vacuum energies unambiguously,
a regularization is extremely useful to define the light-like
quantization surface as a limit of space-like surfaces.
We use space-like quantization surfaces which interpolate between
the ordinary spatial surface and the light-like surface \cite{Horn}.
Light-cone quantization is defined as a limit from
the space-like surface to the light-like one.
This method enabled us to compute the effective potential of light-cone
field theories unambiguously.
As illustrative examples for the effective potential with constant
order parameters, we have studied the Gross-Neveu model,
the $SU(N)$ Thirring model in two-dimensions, and the $O(N)$ vector
models in two, three, and four dimensions using the large $N$ limit.
The previous treatments of these models employed Hartree type
methods and did not compute vacuum energies and effective potentials
 \cite{OhtaThies}.
We find that the vacuum energies are independent of the interpolating
angle to define the quantization surface.
We have performed the renormalization of the effective potential
explicitly \cite{Cole74}.
As an example of the case with nonconstant order parameter,
we have also studied the two-dimensional QCD with quarks in the
fundamental representation.
We explicitly obtain in the large $N$ limit the vacuum energies which
give the gap equation as the stationary point.
The gap equation turns out to depend on the interpolating angle which
defines the quantization surface and the gauge parameter.
In the limit of spatial quantization surface, our gap equation agrees
with the axial gauge result \cite{Li}.

Our results suggest that one can neglect the constraint zero mode
problem once the possible vacuum condensate is determined by our
method of vacuum energy and effective potential.
Although the zero mode fluctuations around the vacuum value give
induced interactions among nonzero modes through the zero mode
constraint, these interaction terms are always multiplied by
inverse powers of the length of the compact spatial dimension
and should disappear as we let the length to go to infinity.
The only subtlety should lie in the determination of the vacuum
condensate, and it can be most efficiently incorporated by means of
effective potential.
Therefore we propose as a practical method that the possible vacuum
condensate be determined by using our vacuum energy and effective
potential and that the induced interactions due to the zero mode
fluctuations should be neglected in using the discretized
light-cone or other approaches to obtain mass spectra and wave
functions.

In sect. 2, we study the Gross-Neveu model and the $SU(N)$ Thirring
model.
In sect. 3, the $O(N)$ $\lambda\phi^4$ model is worked out.
In sect. 4, we compute vacuum energies of QCD.
Our conventions and useful formulas are summarized in appendix.
%
%
\vspace{7mm}
\pagebreak[3]
\addtocounter{section}{1}
\setcounter{equation}{0}
\setcounter{subsection}{1}
\setcounter{footnote}{0}
\begin{center}
{\large {\bf \thesection. Gross-Neveu Model and Its Generalizations }}
\end{center}
\begin{center}
{\bf \thesubsection. Massive Gross-Neveu Model}
\end{center}
\nopagebreak
\medskip
\nopagebreak
\hspace{3mm}

We consider the large $N$ limit of the Gross-Neveu model
which contains a four-fermion interaction
among $N$ component Dirac fields $\psi^a, a=1, \cdots, N$
in two dimensions \cite{GN}
\begin{equation}
{\cal L}=\bar \psi^a\left(i\gamma^{\mu}\partial_{\mu}-m_0\right)\psi^a
+{g_0 \over 2N}\left(\bar\psi^a\psi^a\right)^2 ,
\label{GNLag}
\end{equation}
where $m_0$ and $g_0$ are bare mass and bare coupling constant,
respectively.
This model has a global $U(N)$ symmetry
$ \psi^a\rightarrow U^a{}_b\psi^b $.
Moreover, it possesses a discrete chiral symmetry
when $m_0 = 0$
\begin{equation}
\psi^a \rightarrow \gamma_5 \psi^a.
\label{dischi}
\end{equation}

By introducing an auxiliary field $\sigma$, we can obtain an equivalent
Lagrangian
\begin{equation}
{\cal L}_{\sigma}
=\bar \psi^a\left(i\gamma^{\mu}\partial_{\mu}-\sigma\right)\psi^a
+{Nm_0 \over g_0}\sigma -{N \over 2g_0}\sigma^2,
\label{GNLagax}
\end{equation}
which reduces to the original one (\ref{GNLag}) after integrating
over $\sigma$, since
\begin{equation}
{\cal L}={\cal L}_{\sigma}+{N \over 2g_0}\left(\sigma+{g_0 \over N}
\bar{\psi}^a\psi^a-m_0\right)^2-{Nm_0^2 \over 2g_0}.
\end{equation}

Our goal is to compute vacuum energies in the light-cone
quantization. This procedure, however, encounters ill-defined
quantities if one performs quantization naively on light-like surface.
In order to overcome this problem, we shall define the light-like
quantization surface as a limit from the space-like surface.
This procedure can be regarded as a regularization to define
the singular light-cone quantization properly.
In this way, we can unambiguously compute the vacuum energies.
To this end, we use the coordinate system which interpolates the
light-cone and ordinary coordinates \cite{Horn}
\begin{equation}
\pmatrix{x^+ \cr x^-} =
\pmatrix{\sin {\theta \over 2} & \cos{\theta \over 2}
\cr \cos{\theta \over 2} & -\sin{\theta \over 2}}
\pmatrix{x^0 \cr x^1},
\label{interpolating}
\end{equation}
where $\theta$ is a parameter defined in the region
${\pi \over 2} < \theta \le \pi$.
In this frame, metric tensor becomes
\begin{equation}
g_{\mu\nu}=g^{\mu\nu}=
\pmatrix{c & s \cr s & -c}, \quad
\mu,\nu=+,-, \quad
c\equiv -\cos{\theta}, \ s\equiv \sin{\theta}.
\label{metric}
\end{equation}
In quantization, we regard $x^+$ and $x^-$
as time and space, respectively.
Note that ordinary time quantization corresponds to the limit
\begin{equation}
\theta \rightarrow \pi, c=-\cos{\theta} \rightarrow 1, s=\sin{\theta}
 \rightarrow 0,
\qquad \quad \quad x^+=x^0, \quad x^-=-x^1,
\end{equation}
and light-cone quantization corresponds to the limit
\begin{equation}
\theta \rightarrow {\pi \over 2}, c=-\cos{\theta} \rightarrow 0,
s=\sin{\theta} \rightarrow 1,
\qquad x^+={x^0+x^1 \over \sqrt2}, \ x^-={x^0-x^1 \over \sqrt2},
\end{equation}
Let us emphasize that this change of quantization surface is nothing
to do with the Lorentz transformation.
Conjugate momentum for $\psi^a$ is defined by
\begin{equation}
\pi^a(x)={\partial {\cal L}_{\sigma} \over \partial\partial_+\psi^a(x)}
=i\bar \psi^a\gamma^+ ,
\end{equation}
whose components are given explicitly in
eq.(\ref{spinormomentum}) in appendix.
We can apply the ordinary quantization of Dirac particle
as long as $s \neq 1$ and impose
an anticommutation relation at equal time
\begin{equation}
\left\{ \psi^a_{\alpha}(x), \pi^{b}_{\beta}(y)\right\}_{x^+=y^+}
=
i \delta_{\alpha\beta}\delta^{ab}\delta(x^{-}-y^{-}).
\label{anticom}
\end{equation}
Hamiltonian density is
\begin{equation}
{\cal H}=\pi^a\partial_+\psi^a-{\cal L}_{\sigma}
=\bar \psi^a(-i\gamma^-\partial_-+\sigma)\psi^a-{Nm_0 \over g_0}\sigma
+{N \over 2g_0}\sigma^2.
\label{Hamil}
\end{equation}

\begin{figure}[htbp]
    \begin{center}
     \leavevmode
       \epsfxsize = 14.5cm
\hspace{.5mm}
       \epsfbox{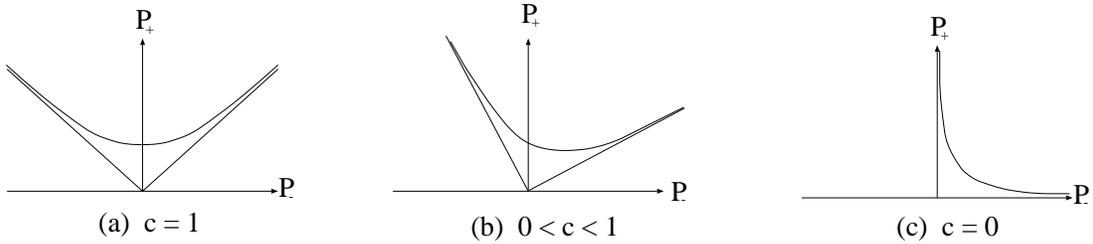}
    \end{center}
\caption{
Dispersion relations of a free massive particle on (a) usual,
  (b) an interpolating and (c) light-cone quantization surfaces.}
\label{dispersion}
\end{figure}

In the large $N$ limit, vacuum energy is given by the fermion one loop
contributions.
Therefore we shall treat the auxiliary field $\sigma$ as a background
field.
Since we are interested in the effective potential to determine the
vacuum expectation value, we take $\sigma$ as a constant
background.
There exist only quadratic terms in the quantum field $\psi^a$ in the
Lagrangian.
By solving the equation of motion for $\psi^a$, we obtain the
light-cone energy
\begin{equation}
p_+={1 \over c}\left(-sp_-+\omega_p\right), \quad
\omega_p=\sqrt{(p_-)^2+c\sigma^2}.
\label{p+}
\end{equation}
and the corresponding spinor is given in eq.(\ref{freespinor})
in appendix.
If we take the light-like limit
$c=-\cos\theta \rightarrow 0, s=\sin\theta \rightarrow 1$,
we obtain finite energy only for positive momenta $p_- > 0$ as shown
in Fig.\ref{dispersion}.
\begin{equation}
\lim_{c\rightarrow 0}p_+=
{\sigma \over 2p_-}
{}.
\label{lightconedispersion}
\end{equation}
%
%

To avoid possible infrared divergences we compactify  $x^{-}$
direction and impose a periodic boundary condition
$\psi^a(x^-)=\psi^a(x^-+2L)$,
which gives discrete momenta
\begin{equation}
p_{n-}={\pi n \over L}, \quad n \in  {\bf Z}.
\label{discretep-}
\end{equation}
Other boundary conditions are also allowed.
Therefore the fermion field operator is expanded into modes
\begin{equation}
\psi^a(x)={1 \over \sqrt{2L}}
\sum_n \left[{\rm e}^{-ip_n x} u(p_n)b_n^a +
{\rm e}^{ip_n x} v(p_n)d_n^{a\dagger} \right],
\label{dismodeex}
\end{equation}
where $p_n=\left(p_{n+}, p_{n-}\right)$ is defined in eq.
(\ref{p+}) and (\ref{discretep-}).
Anticommutation relation (\ref{anticom}) becomes
\begin{equation}
\{b_n^a, b_m^{b\dagger}\}=
\delta_{n,m}\delta^{ab},
\quad
\{d_n^a, d_m^{b\dagger}\}=
\delta_{n,m}\delta^{ab}.
\end{equation}
The Hamiltonian is given by using (\ref{Hamil}) and  (\ref{dismodeex})
\begin{equation}
P_+=\int_{-L}^Ldx^- {\cal H}(x)
=\sum_n p_{n+}\left(b_n^{a\dagger}b_n^a-d_n^a d_n^{a\dagger}\right)
+ 2L \left(-{Nm_0 \over g_0}\sigma+{N \over 2g_0}\sigma^2\right).
\end{equation}
Since the vacuum of this Hamiltonian is the Fock vacuum satisfying
$
b_n^a\left\vert 0 \right\rangle=
d_n^a\left\vert 0 \right\rangle=0,
$
we obtain the vacuum energy density
in the leading order of the $1/N$ expansion
\begin{equation}
{1 \over 2L}
\left\langle 0 \right\vert P_+ \left\vert 0 \right\rangle
=V(\sigma),
\end{equation}
\begin{equation}
V(\sigma)=V_0-{Nm_0 \over g_0}\sigma+{N \over 2g_0}\sigma^2
+V_{1-{\rm loop}}(\sigma),
\end{equation}
\begin{equation}
V_{1-{\rm loop}}(\sigma)
=
-{N \over 2L}\sum_n p_{n+}
=
-{N \over 2L}\sum_n {1 \over c}
\left[-s{n\pi \over L}
+\sqrt{\left({n\pi \over L}\right)^2+c\sigma^2}\right] .
\label{disvacenrg}
\end{equation}
We have introduced a constant $V_0$ to renormalize the cosmological
constant.
We observe that $V_{1-{\rm loop}}$ appears to depend on the
parameter $c=-\cos \theta, s=\sin \theta$ in eq.(\ref{interpolating})
to define the quantization surface.
Since one-loop vacuum energy density has no infrared divergence,
we can now take $L\rightarrow \infty$ limit and obtain
\begin{equation}
V_{1-{\rm loop}}(\sigma)
=
-N \int {dp_- \over 2\pi}
\left[{-sp_-+\sqrt{(p_-)^2+c\sigma^2} \over c} \right].
\label{vacenrgint}
\end{equation}

Since the vacuum energy density is UV divergent,
we apply the Pauli-Villars regularization
\begin{equation}
V_{1-{\rm loop}}^{PV}(\sigma)
=\lim_{\Lambda_i\rightarrow \infty}
\left[V_{1-{\rm loop}}(\sigma)
-
\sum_i a_i V_{1-{\rm loop}}(\Lambda_i)
\right].
\end{equation}
By requiring
$\sum_i a_i=1, \quad \sum_ia_i \Lambda_i^2=\sigma^2$,
we obtain
\begin{equation}
V_{1-{\rm loop}}^{PV}(\sigma)
=-{N \over 4\pi}
\lim_{\Lambda_i\rightarrow \infty}
\left(
\sum_i a_i \Lambda_i^2 \log \Lambda_i^2 - \sigma^2 \log \sigma^2
\right)  .
\label{pvvacenrgy}
\end{equation}
We find that the positive ($p_->0$) and negative ($p_-<0$) momentum
region contribute equally to the one-loop effective potential.
It is interesting to observe that the vacuum energy no longer
depends on the parameter $c=-\cos \theta, s=\sin \theta$ in
eq.(\ref{pvvacenrgy}) in contrast to the expression
(\ref{vacenrgint}) before the UV regularization and the
momentum integration.
The result is identical to that obtained by the usual covariant
calculations.
On the other hand,
if we take the limit of light-like quantization surface
$c=-\cos \theta \rightarrow 0$ before integrating over the momentum
$p_-$, we find that only the positive momentum region contributes
as seen from eq.(\ref{lightconedispersion}).
The integral becomes infrared divergent.
Even if we regularize the integral by introducing the compact spatial
dimension, the contribution of zero mode is still ambiguous and
the nonzero mode alone gives a result different from the above.
As Fig.\ref{dispersion} suggests, this situation arises since
contributions from
$p_-<0$ region are squeezed into the infrared divergent zero mode
contribution and become ambiguous if the light-cone limit is
taken inside the momentum integral.
Therefore we conclude that the light-cone limit does not commute with
the momentum integration and that the light-cone limit must be
taken after integrating over the momentum.
To test the sensitivity of the procedure to define the
light-like surface as a limit of the spacelike surface,
we have also computed the vacuum energy using another
choice of the limit from the spacelike surface \cite{Lenz} with
$x^+={1 \over \sqrt2}\left[\left(1+{\epsilon \over 2L}\right)x^0
+\left(1-{\epsilon \over 2L}\right)x^1\right]$ and
$x^-={1 \over \sqrt2}\left[x^0-x^1\right]$, and have found the identical
result provided we perform momentum integration before taking the
limit.

A simple choice of the Pauli-Villars regulators
$a_1=2, \quad a_2=-1, \quad \Lambda_2^2=2\Lambda_1^2-\sigma^2$
gives
\begin{equation}
V_{1-{\rm loop}}^{PV}(\sigma)
=-{N \over 4\pi}
\lim_{\Lambda_1\rightarrow \infty}
\left[
\sigma^2 \left(\log {2\Lambda_1^2 \over 2\sigma^2}+1\right)
-2 \Lambda_1^2 \log 2
\right].
\end{equation}
We impose renormalization conditions for
the cosmological constant, the mass, and the coupling constant
\begin{equation}
V(\sigma=0)=0, \quad
{\partial V \over \partial \sigma}\bigl(\sigma=0\bigr)=-{Nm \over g},
\quad
{\partial^2 V \over \partial \sigma^2}\bigl(\sigma=\mu\bigr)={N \over g},
\end{equation}
where $m$ and $g$ are the renormalized mass and the renormalized
coupling constant, respectively.
The renormalized effective potential is finally given by
\begin{equation}
V(\sigma)=-{Nm \over g}\sigma+{N \over 2g}\sigma^2
+{N \over 4\pi}\left[
\sigma^2 \left(\log {\sigma^2 \over \mu^2}+1\right)
\right].
\end{equation}
Minimizing the effective potential, we obtain a nonvanishing vacuum
expectation value for $\sigma$
\begin{equation}
\sigma \approx {m \over |m|} \mu {\rm e}^{-1-{\pi \over g}}
+{\pi m \over g},
\end{equation}
which implies the spontaneous breakdown of the discrete chiral
symmetry (\ref{dischi}).
As was discussed in \cite{GN}, it follows that the
chiral condensate occurs
$\langle \bar{\psi}^a\psi^a \rangle \neq 0$ .
%
\vspace{7mm}
\pagebreak[3]
\addtocounter{section}{0}
\setcounter{equation}{27}
\setcounter{subsection}{2}
\setcounter{footnote}{0}
\begin{center}
{\bf \thesubsection. $SU(N)$ Thirring Model }
\end{center}
\nopagebreak
\medskip
\nopagebreak
\hspace{3mm}

The $SU(N)$ Thirring model is a generalization of the Gross-Neveu model
\cite{Witten}
\begin{equation}
{\cal L}=\bar \psi^a\left(i\gamma^{\mu}\partial_{\mu}-m_0\right)\psi^a
+{g_0 \over 2N}\left[\left(\bar\psi^a\psi^a\right)^2-
\left(\bar\psi^a\gamma_5\psi^a\right)^2\right] .
\end{equation}
Using auxiliary fields $\sigma$ and $\pi$ corresponding to
the scalar and pseudoscalar fermion bilinears, we obtain the equivalent
Lagrangian
\begin{equation}
{\cal L}_{\pi \sigma}
=\bar \psi^a\left(i\gamma^{\mu}\partial_{\mu}
-\sigma-i\pi\gamma_5\right)\psi^a
+{Nm_0 \over g_0}\sigma -{N \over 2g_0}\left(\sigma^2+\pi^2\right).
\end{equation}
This model has a global $U(N)$ symmetry
$\psi^a\rightarrow U^a{}_b\psi^b$.
It also has a continuous chiral symmetry when $m_0=0$
\begin{equation}
\psi^a\rightarrow {\rm e}^{i\beta\gamma_5}\psi^a, \quad
\sigma+i\pi\rightarrow {\rm e}^{-2i\beta}
\left(\sigma+i\pi\right).
\end{equation}
Following the same procedure as that of the Gross-Neveu model,
we obtain the renormalized effective potential in the large $N$ limit
\begin{equation}
V(\sigma, \pi)=-{Nm \over g}\sigma+{N \over 2g}
\left(\sigma^2+\pi^2\right)
+{N \over 4\pi}\left[
\left(\sigma^2+\pi^2\right)
\left(\log {\sigma^2+\pi^2 \over \mu^2}+1\right)
\right],
\end{equation}
which depends on two
spacetime-independent background fields $\sigma$ and $\pi$.

{}By minimizing this effective potential, one finds that the $\sigma$
acquires a nonvanishing vacuum expectation value and
the continuous chiral symmetry is spontaneously broken in
two dimensions.
It has been observed \cite{Witten} that higher order contributions
introduces power law decay for the correlation function
$\langle\bar{\psi}^a(1-\gamma^5)\psi^a(x)\bar{\psi}^b(1+\gamma^5)
\psi^b(y)\rangle \propto |x-y|^{-{1 \over N}}$
as $|x-y|\rightarrow\infty$.
This Berezinski-Kosterliz-Thouless type behavior \cite{KT} makes
the correlation
function compatible with the Coleman's theorem \cite{Cole73}.
Since this behavior shows that the chiral symmetry is almost broken,
the leading order result gives physically correct picture
of the Gross-Neveu model.
%
%
\vspace{7mm}
\pagebreak[3]
\addtocounter{section}{1}
\setcounter{equation}{0}
\setcounter{subsection}{0}
\setcounter{footnote}{0}
\begin{center}
{\large {\bf \thesection. $O(N)$ $\lambda \phi^4$ Model in $d$-Dimensions }}
\end{center}
\nopagebreak
\medskip
\nopagebreak
\hspace{3mm}

In this section, we consider $N$ component scalar field $\phi^a$
($a=1, \cdots, N$) with the $O(N)$ invariant quartic interaction
in $d$-dimensions ($d=2, 3, 4$)
\begin{equation}
{\cal L}={1 \over 2}\partial_{\mu} \phi^a\partial^{\mu}\phi^a
-{m_0^2 \over 2}\phi^a\phi^a
-{g_0 \over 8N}\left(\phi^a\phi^a\right)^2.
\end{equation}
This theory is invariant under global $O(N)$ transformations
$\phi^a\rightarrow U^a{}_b\phi^b$.
We shall show that the vacuum energy of this
model can be defined in the light-cone quantization.
Introducing an auxiliary field $\sigma$, we obtain an equivalent
Lagrangian
\begin{equation}
{\cal L}_{\sigma}
=
{1 \over 2}\partial_{\mu} \phi^a\partial^{\mu}\phi^a
-{1 \over 2}\sigma \phi^a\phi^a
-{Nm_0^2 \over g_0}\sigma +{N \over 2g_0}\sigma^2.
\end{equation}

In the interpolating coordinates (\ref{interpolating}), the Lagrangian
becomes
\begin{equation}
{\cal L}_{\sigma}
=
{c \over 2}\left[\left(\partial_{+} \phi^a\right)^2
-\left(\partial^{-}\phi^a\right)^2\right]
+s\partial_+\phi^a\partial_-\phi^a
-{1 \over 2}(\partial_{\bot} \phi^a)^2
-{1 \over 2}\sigma \phi^a\phi^a
-{Nm_0^2 \over g_0}\sigma +{N \over 2g_0}\sigma^2.
\end{equation}
Conjugate momentum for $\phi^a$ is defined by
\begin{equation}
\pi^a={\partial {\cal L}_{\sigma} \over \partial\partial_+\phi^a}
=c\partial_+\phi^a+s\partial_-\phi^a .
\end{equation}
Regarding the auxiliary field $\sigma$ as spacetime-independent
background field as before, this model reduces to free massive bosons.
To avoid possible infrared divergences,
we compactify $x^-$,$x^{\bot}$ directions
and impose periodic boundary conditions 
$
\phi^a(x^-,x^{\bot})=\phi^a(x^-+2L,x^{\bot})=\phi^a(x^-,x^{\bot}+2L),
$
where $x^{\bot}=x^i, \ \ i=2,3, \cdots, d-1$.
As a result, momenta take discrete values
$
p_-={\pi n \over L}, \quad
 p_{\bot}={\pi  n \over L}.
$
Solving the equation of motion, $\phi^a$ can be expanded into
modes
\begin{equation}
\phi^a(x)={1 \over \sqrt{2L}}
\sum_n {1 \over \sqrt{2\omega_n}}
\left[{\rm e}^{-ip \cdot x} a^a( n) +
{\rm e}^{ip \cdot x} a^{a\dagger}( n) \right],
\label{modephi}
\end{equation}
where $p \cdot x = p_{+}x^{+}+p_{-}x^{-}+p_{\bot}x^{\bot}$ with
$p_{+}=(-sp_-+\omega_p)/c, \omega_p=\sqrt{(p_-)^2+(p^{\bot})^2+c\sigma}$.
Imposing the canonical commutation relation at $x^+=y^+$,
we obtain
\begin{equation}
\left[a^a( n), a^{a\dagger}( m)\right] =
\delta_{n, m}\delta^{ab}.
\end{equation}

Using (\ref{modephi}), the Hamiltonian is given by
\begin{eqnarray}
P_+
 &\!\!\!=&\!\!\!
\int_{-L}^Ld^{d-1} x
\left[\pi^a\partial_+\phi^a-{\cal L}_{\sigma}\right]
\nonumber \\
&\!\!\!=&\!\!\!
\sum_n{1 \over 2}\left[ a^{a\dagger}( n)a^a( n)+
 a^a( n) a^{a\dagger}( n)\right]
+(2L)^{d-1}\left({Nm_0^2 \over g_0}\sigma -{N \over 2g_0}\sigma^2
\right) .
\end{eqnarray}
In the large $N$ limit, vacuum is the Fock vacuum
$\left\vert 0 \right\rangle$ and the
vacuum energy density is given by
\begin{equation}
V(\sigma)
=
{1 \over (2L)^{d-1}}
\left\langle 0 \right\vert P_+ \left\vert 0 \right\rangle,
\label{vacengy}
\end{equation}
\begin{equation}
V(\sigma)=V_0+{Nm_0^2 \over g_0}\sigma-{N \over 2g_0}\sigma^2
+V_{1-{\rm loop}}(\sigma),
\end{equation}
\begin{equation}
V_{1-{\rm loop}}(\sigma)
=
{N \over (2L)^{d-1}}{1 \over 2}\sum_{ n}
\left[{-sp_-+\sqrt{(p_-)^2+c((p^{\bot})^2+\sigma)} \over c} \right] .
\label{oneloop}
\end{equation}
Since there is no infrared
singularity in the vacuum energy density,
we can take $L\rightarrow \infty$ by replacing
the discrete sum in (\ref{oneloop}) by a momentum
integration.
Similarly to the case of the Gross-Neveu model, the above expression for
the one-loop vacuum energy appears to depend on the parameter
$c=-\cos \theta, s=\sin \theta$ in eq.(\ref{interpolating})
to define the interpolating quantization surface.
In the following, however, we shall work out
explicit forms of the effective potential
in the case of $d=2,3,4,$
and shall find the result to be independent of
the parameter $c=-\cos\theta, s=\sin \theta$ and to agree with those
given by the covariant formalism.

In order to define the effective potential as a function of the
constant classical field corresponding to the $O(N)$ vector field
$\phi^a$,
we introduce the spacetime-independent source $J^a$ coupled to
$\phi^a$
\begin{equation}
{\cal L}_{\sigma J^a}
=
{\cal L}_{\sigma}+J^a\phi^a
=
-{\sigma \over 2}\left(\phi^a-{J^a \over\sigma}\right)^2
-{Nm_0^2 \over g_0}\sigma +{N \over 2g_0}\sigma^2
+{J^aJ^a \over 2\sigma}.
\end{equation}
The generating function $W[J^a, \sigma]$
and the classical field $\varphi^a$ are defined by
\begin{equation}
W(J^a, \sigma)\equiv
-{1 \over (2L)^{d-1}}
\left\langle 0 \right\vert P_+ \left\vert 0 \right\rangle
=
-\left({Nm_0^2 \over g_0}\sigma -{N \over 2g_0}\sigma^2
+V_{1-{\rm loop}}(\sigma)\right)
+{J^aJ^a \over 2\sigma},
\label{generating}
\end{equation}
\begin{equation}
\varphi^a\equiv {\partial W \over \partial J^a} = {J^a \over \sigma}.
\end{equation}
Performing the Legendre transformation, we obtain the effective
potential
\begin{equation}
V(\varphi^a, \sigma)
=J^a\varphi^a-W(J^a, \sigma)
={1 \over 2}\sigma\varphi^a\varphi^a+
{Nm_0^2 \over g_0}\sigma -{N \over 2g_0}\sigma^2
+V_{1-{\rm loop}}(\sigma),
\end{equation}
which depends on the classical field $\varphi^a$ and the background
field $\sigma$.

As can be seen easily, the effective potential is UV divergent and
the degree of divergence depends on the dimensions of spacetime.
To regularize the UV divergence, we use the Pauli-Villars
regularization method
\begin{equation}
V_{1-{\rm loop}}^{PV}(\sigma)
=\lim_{\Lambda_i\rightarrow \infty}
\left[V_{1-{\rm loop}}(\sigma)
-
\sum_i a_i V_{1-{\rm loop}}(\Lambda_i)
\right].
\end{equation}

We need to renormalize the model in each dimensions separately.
In two dimensions, it is easy to see that the one loop contribution
to the effective potential
is equivalent to that of the Gross-Neveu model up to a constant factor
\begin{equation}
V_{1-loop}^{O(N)}(\sigma)=
-{1 \over 2}V_{1-loop}^{Gross-Neveu}(\sqrt{\sigma}).
\end{equation}
We need to renormalize the cosmological constant and the mass
but not the coupling constant
\begin{equation}
V(\sigma=0)=0, \qquad
{\partial V \over \partial \sigma}\bigl(\sigma=\mu^2\bigr)
={N(m^2-\mu^2) \over g}.
\end{equation}
The renormalized effective potential in two dimensions is given by
\begin{equation}
V(\varphi^a, \sigma)={1 \over 2}\sigma\varphi^a\varphi^a+
{Nm^2 \over g}\sigma-{N \over 2g}\sigma^2
+{N \over 8\pi}
\sigma \left(\log {\mu^2 \over \sigma}+1\right).
\end{equation}
Searching for the stationary point with respect to $\sigma$
\begin{equation}
0={\partial V \over \partial\sigma}\bigl(\varphi^a, \sigma\bigr)
=
{1 \over 2}\varphi^a\varphi^a+
{Nm^2 \over g}-{N \over g}\sigma
+{N \over 8\pi}\log {\mu^2 \over \sigma},
\end{equation}
the effective potential depending only on $\varphi^a$ is given by
\begin{equation}
V(\varphi^a)
\equiv V(\varphi^a,\sigma=\sigma(\varphi^a))
=
{N \over 2g}\left[\sigma(\varphi^a)\right]^2.
\end{equation}
One finds that $\varphi^a$ vanishes at the minimum
and the $O(N)$ symmetry is not broken.
This result is consistent with the Coleman's theorem and the leading
order approximation of the $1/N$ expansion yields a reliable
result \cite{Cole74}.

In three dimensions, the one loop contribution after choosing the
Pauli-Villars regulators is
\begin{equation}
V_{1-loop}^{PV}(\sigma)=
-{N \over 12\pi}\left(\sigma^{3 \over 2}-\sum_ia_i\Lambda_i^3\right).
\end{equation}
We need to renormalize the cosmological constant and the mass
but not the coupling constant
\begin{equation}
V(\sigma=0)=0, \qquad
{\partial V \over \partial \sigma}\bigl(\sigma=0\bigr)
={Nm^2 \over g}.
\end{equation}
The renormalized effective potential is thus given by
\begin{equation}
V(\varphi^a, \sigma)={1 \over 2}\sigma\varphi^a\varphi^a+
{Nm^2 \over g}\sigma-{N \over 2g}\sigma^2
-{N \over 12\pi}
\sigma^{3 \over 2}.
\end{equation}
Expressing $\sigma$ by solving the stationarity condition
$\partial V/\partial \sigma=0$,
we obtain the effective potential
$V(\varphi^a)\equiv V(\varphi^a,\sigma=\sigma(\varphi^a))$
depending only on the classical field $\varphi^a$ \cite{Cole74}.

In four dimensions, the regularized effective potential is given by
\begin{equation}
V_{1-loop}^{PV}(\sigma)=
{N \over 32\pi}\left(\sigma^2\log \sigma
-\sum_ia_i\Lambda_i^4\log \Lambda_i^2\right).
\end{equation}
In contrast to two and three dimensions, the effective potential
in four dimensions requires renormalization of the cosmological
constant, mass and the coupling constant
\begin{equation}
V(\sigma=0)=0, \quad
{\partial V \over \partial \sigma}\bigl(\sigma=0\bigr)
={Nm^2 \over g}, \quad
{\partial^2 V \over \partial \sigma^2}\bigl(\sigma=\mu^2\bigr)
=-{N \over g}.
\end{equation}
These conditions give the renormalized effective potential
\begin{equation}
V(\varphi^a, \sigma)={1 \over 2}\sigma\varphi^a\varphi^a+
{Nm^2 \over g}\sigma-{N \over 2g}\sigma^2
+{N \over 32\pi}
\sigma^2\left[\log{\sigma \over \mu^2}-{3 \over 2}\right] .
\end{equation}
The background field $\sigma$ is determined as a function of the
classical field $\varphi^a$ by the stationary condition
$\partial V/\partial \sigma=0$.
Eliminating $\sigma$ one finally obtains the effective potential
$V(\varphi^a)\equiv V(\varphi^a,\sigma=\sigma(\varphi^a))$
whose physical meaning is discussed
in detail in \cite{Cole74}, \cite{Abbott}.
%
%
\vspace{7mm}
\pagebreak[3]
\addtocounter{section}{1}
\setcounter{equation}{0}
\setcounter{subsection}{0}
\setcounter{footnote}{0}
\begin{center}
{\large {\bf \thesection. QCD in Two Dimensions in the Large $N$ Limit }}
\end{center}
\nopagebreak
\medskip
\nopagebreak
\hspace{3mm}

QCD in two dimensions is an
another interesting model which exhibits the nontrivial vacuum structure,
namely the quark-antiquark condensation in the large $N$ limit
\cite{Zhitni}--\cite{Li}.
Similarly to the $SU(N)$ Thirring model, higher order corrections
in $1/N$ expansion should introduce the power law decay for the
correlation function of
$<\bar\psi\psi(x)\;\bar\psi\psi(y)> \sim |x-y|^{-\frac{1}{N}}$
as $|x-y|\rightarrow\infty$, in conformity with the Coleman's theorem
\cite{Witten}.
Since power law decay is much milder than the usual exponential
decay, the spontaneous breaking is almost realized and
the leading order in the $1/N$ expansion gives physically sensible
result.

The Lagrangian consists of $SU(N)$ gauge fields $A_{\mu}^a$ and
the quark $\psi^i$ in the fundamental representation
\begin{equation}
{\cal L}=\bar \psi\left(i\gamma^{\mu}D_{\mu}-m\right)\psi
-{1 \over 4}F_{\mu\nu}^a F^{a\mu\nu} ,
\end{equation}
where the field strength $F_{\mu\nu}$ and the
covariant derivative $D_{\mu}$ are defined as
$$
D_{\mu}=\partial_{\mu}+igA_{\mu}, \qquad A_{\mu}=A_{\mu}^a T^a,
$$
$$
F_{\mu\nu}=\partial_{\mu}A_{\nu}-\partial_{\nu}A_{\mu}
+ig[A_{\mu}, A_{\nu}],
\qquad
{\rm tr}\left(T^a T^b\right)={1 \over 2}\delta^{ab}.
$$

We adopt the light-cone gauge
\begin{equation}
A_-=0.
\end{equation}
As an advantage of the light-cone gauge, the remaining gauge field
becomes a dependent variable.
We can eliminate $A_+^a$ by the Gauss law constraint to obtain
the Lagrangian
\begin{equation}
A_+^a
=-{g \over \partial_-^2}\bar \psi\gamma^+T^a\psi
=-{g \over \partial_-^2}J^{+a} ,
\end{equation}
\begin{equation}
{\cal L}=\bar \psi\left(i\gamma^{\mu}\partial_{\mu}-m\right)\psi
+{1 \over 2}J^{+a}{1 \over \partial_-^2} J^{+a} .
\end{equation}
Thus, the Hamiltonian density with $x^+$ as time is given by
\begin{equation}
{\cal H}
=\bar \psi^a(-i\gamma^-\partial_-+m)\psi^a
-{1 \over 2}J^{+a}{1 \over \partial_-^2} J^{+a} .
\end{equation}


To find the vacuum state, let us minimizing the expectation value of the
Hamiltonian over a trial vacuum state.
To exhibit a quark-antiquark condensation,
we choose the trial vacuum state $\left\vert {\Phi} \right\rangle$
which is obtained by the
Bogoliubov transformation from the Fock vacuum state
$\left\vert 0 \right\rangle$ \cite{AD}
\begin{equation}
\left\vert {\Phi} \right\rangle
={1 \over \sqrt{1+\Phi^2(p_-)}}
\Pi_{p_-}\left[1-\Phi(p_-)b^{i\dagger}(p_-)d^{i\dagger}(-p_-)\right]
\left\vert { 0 } \right\rangle ,
\end{equation}
where $b^{i\dagger}(p_-)$ and $d^{i\dagger}(p_-)$ are creation
operators for quark and antiquark with momentum $p_-$, color index $i$.
The state is annihilated by operators given by the Bogoliubov
transformation
\begin{equation}
B^i(p_-)\left\vert {\Phi} \right\rangle
=D^i(p_-)\left\vert {\Phi} \right\rangle
=0 ,
\end{equation}
\begin{eqnarray}
B^i(p_-) &=& {1 \over \sqrt{1+\Phi^2(p_-)}}
\left[b^i(p_-)+\Phi(p_-)d^{i\dagger}(-p_-)\right],
\nonumber \\
D^{i\dagger}(-p_-) &=& {1 \over \sqrt{1+\Phi^2(p_-)}}
\left[d^{i\dagger}(-p_-)-\Phi(p_-)b^i(p_-)\right].
\label{bogoliubovtranfop}
\end{eqnarray}
The quark field can be expressed in terms of the original and
transformed operators
\begin{eqnarray}
\psi^i(x^-) &=&
\int {dp_- \over \sqrt{2\pi}} {\rm e}^{-ip_-x^-}
\left[u(p_-)b^i(p_-) + v(-p_-)d^{i\dagger}(-p_-)\right]
\nonumber \\
&=&
\int {dp_- \over \sqrt{2\pi}} {\rm e}^{-ip_-x^-}
\left[U(p_-)B^i(p_-)+V(-p_-)D^{i\dagger}(-p_-)\right] ,
\label{bogoliubov}
\end{eqnarray}
where $u(p_-)$ and $v(p_-)$ are original free massive spinors in
eq.(\ref{freespinor}), while $U(p_-)$ and $V(p_-)$ are transformed
ones in eq.(\ref{bogoluibovspinor}) in appendix.
The commutator of quark fields at equal time can be parametrized
by the order parameter $\Phi(p_-)$ of quark-antiquark condensation
\begin{eqnarray}
&& \left\langle {\Phi} \right\vert
{1 \over 2}\left[\psi_{\alpha}^i(0,x^-),
\bar \psi^{\beta j}(0,y^-)\right]
\left\vert {\Phi} \right\rangle
\nonumber \\
&=& {1 \over 2}\delta^{i j} \int {dp \over 2\pi} {\rm e}^{-ip(x^--y^-)}
\left[\left({1-\Phi^2(p) \over 1+\Phi^2(p)}{m \over \omega_p}
-{2\Phi(p) \over 1+\Phi^2(p)}{p \over \sqrt{c}\omega_p}\right)
\right.
\nonumber \\
&& \left.-\left({1-\Phi^2(p) \over 1+\Phi^2(p)}{p \over c\omega_p}
+{2\Phi(p) \over 1+\Phi^2(p)}{m \over \sqrt{c}\omega_p}\right)\gamma_-
\right]_{\alpha}{}^{\beta}.
\end{eqnarray}
We abbreviate the light-cone momenta $p_-, q_-$ as $p, q$ henceforth.

Vacuum energy density is given by
\begin{eqnarray}
\left\langle {\Phi} \right\vert
{\cal H}
\left\vert {\Phi} \right\rangle_{\rm ren}
&\!\!\!
\equiv
&\!\!\!
 \left\langle {\Phi} \right\vert
{\cal H}
\left\vert {\Phi} \right\rangle -
 \left\langle 0 \right\vert
{\cal H}
\left\vert 0 \right\rangle
= \int{dp \over 2\pi} p_+ N {2\Phi^2(p) \over 1+\Phi^2(p)}
\nonumber \\
&\!\!\!
 +
&\!\!\!
\int{dpdq \over (2\pi)^2(p-q)^2}
{g^2(N^2-1) \over 4 \omega_q\omega_p}
{1 \over 1+\Phi^2(q)}
{1 \over 1+\Phi^2(p)}
\nonumber \\
&\!\!\!
&\!\!\!
 \left[
\left(\omega_q\omega_p+qp+cm^2\right)
\left(\Phi(q)-\Phi(p)\right)^2
\right.
+\left(\omega_q\omega_p-qp-cm^2\right)
\left(1+\Phi(q)\Phi(p)\right)^2
\nonumber \\
&\!\!\!
&\!\!\!
 \left.+2\sqrt{c}m(q-p)
\left(\Phi(q)-\Phi(p)\right)
\left(1+\Phi(q)\Phi(p)\right)
\right] ,
\end{eqnarray}
where the light-cone energy is given as
\begin{equation}
p_+={-sp_-+\omega_p\over c}, \qquad \omega_p=\sqrt{(p_-)^2+cm^2} .
\label{freefermionenergy}
\end{equation}
We have subtracted the vacuum energy of the Fock vacuum to obtain the
renormalized vacuum energy
$\left\langle {\Phi} \right\vert
{\cal H}
\left\vert {\Phi} \right\rangle_{\rm ren}
$ .
%
The order parameter $\Phi(p_-)$ is determined by the extremum
condition
\begin{eqnarray}
0&\!\!\!=&\!\!\! {\delta
\left\langle {\Phi} \right\vert
{\cal H}
\left\vert {\Phi} \right\rangle_{\rm ren} \over
\delta \Phi(p)}
=
{2N \over \pi (1+\Phi^2(p))^2}
\Biggl[
p_+\Phi(p)
\nonumber \\
&& -{g^2(N^2-1) \over 8\pi N}
\int {dq \over (p-q)^2}
{1 \over 1+\Phi^2(q)}{1 \over \omega_q\omega_p}
\nonumber \\
&& \left\{2(qp+cm^2)\left(1+\Phi(q)\Phi(p)\right)
\left(\Phi(q)-\Phi(p)\right)
\right.
\nonumber \\
&& \left.
+\sqrt{c}m(q-p)\left(
\left(1+\Phi(q)\Phi(p)\right)^2-
\left(\Phi(q)-\Phi(p)\right)^2
\right)
\right\}
\Biggr] .
\label{gapeq}
\end{eqnarray}
This equation is the gap equation in the gauge $A_-=0$.
The gap equation in the axial gauge ($c=1, A_1=0$) for massless
$\mbox{QCD}_2$ is given before \cite{BG}, \cite{Li}.
The gap equation (\ref{gapeq}) in our gauge $A_-=0$ depends
on the parameter $c=-\cos\theta, s=\sin\theta$ defining
the interpolating quantization surface.
Even in the case of $m=0$, the first term of our gap equation
contains the factor $p_+$ which is asymmetric in
$p_- \leftrightarrow -p_-$ as seen in eq.(\ref{freefermionenergy}).
Therefore it will give solutions $\Phi(p_-)$ asymmetric in
$p_- \leftrightarrow -p_-$ which
is different from the axial gauge solution.
We hope that this gauge dependence should disappear when we compute
gauge invariant quantities like the chiral condensate
\begin{equation}
\left\langle {\Phi} \right\vert
\bar \psi(x) \psi(x)
\left\vert {\Phi} \right\rangle
= N \int_{-\infty}^{\infty} {dp_- \over 2\pi}\left[
{2\Phi(p_-) \over 1+\Phi^2(p_-)}{p_- \over \sqrt{c}\omega_p}
-{1-\Phi^2(p_-) \over 1+\Phi^2(p_-)}{m \over \omega_p}\right] .
\end{equation}
A partial numerical evidence for this gauge independence has been given
already \cite{Zhitni}--\cite{Li}.
\par

It is an interesting problem to study two-dimensional QCD with matter
in adjoint representation such as supersymmetric QCD \cite{MaSaSa}.
We are looking for more powerful methods than 1/N expansion.

\vspace{5mm}
%
%
This work is supported in part by Grant-in-Aid for
Scientific Research (S.K.) and (No.05640334) (N.S.)
from the Ministry of Education, Science and Culture.
%
%
\def\numberbysectiona{\@addtoreset{equation}{section}
\def\theequation{A.\arabic{equation}}}
\numberbysectiona
\vspace{7mm}
\pagebreak[3]
\setcounter{section}{1}
\setcounter{equation}{0}
\setcounter{subsection}{0}
\setcounter{footnote}{0}
\begin{center}
{\large{\bf Appendix }}
\end{center}
\nopagebreak
\medskip
\nopagebreak
\hspace{3mm}

We summarize our notations and useful formulas.
Our $\gamma$ matrices in two dimensions are
\begin{eqnarray}
\gamma^0 &=& \sigma_2 = \pmatrix{0 & -i \cr i & 0},
\qquad \gamma^1 =i\sigma_1 = \pmatrix{0 & i \cr i & 0},
\nonumber \\
\gamma_5 &=& \gamma^0\gamma^1=\sigma_3 = \pmatrix{1 & 0 \cr 0 & -1},
\nonumber \\
\gamma^+ &=& \gamma^0 \sin{\theta \over 2} + \gamma^1 \cos{\theta \over 2}
= \pmatrix{0 & -i\sqrt{1-s} \cr i\sqrt{1+s} & 0},
\nonumber \\
\gamma^- &=& \gamma^0 \sin{\theta \over 2} - \gamma^1 \cos{\theta \over 2}
= \pmatrix{0 & -i\sqrt{1+s} \cr -i\sqrt{1-s} & 0}.
\end{eqnarray}
The fermion $\psi$ is a two component Dirac spinor
\begin{equation}
\psi=
\pmatrix{\psi_R \cr \psi_L}.
\end{equation}
The Lagrangian of the free massive fermion and the conjugate
momentum $\pi$ for $\psi$ are
\begin{equation}
{\cal L}_{m}=
\pmatrix{\psi_R^{*} & \psi_L^*}
\pmatrix{i(\sqrt{1+s}\partial_+-\sqrt{1-s}\partial_-) & i m
\cr -i m & i(\sqrt{1-s}\partial_+-\sqrt{1+s}\partial_-)}
\pmatrix{\psi_R \cr \psi_L} ,
\end{equation}
\begin{equation}
\pi(x)={\partial {\cal L}_{\sigma} \over \partial\partial_+\psi(x)}
=i\bar \psi\gamma^+
=
\pmatrix{\psi_R^{*} & \psi_L^{*}}
\pmatrix{i\sqrt{1+s} & 0
\cr 0 & i\sqrt{1-s}}.
\label{spinormomentum}
\end{equation}
%
Spinors $u(p_{-})$ and $v(p_{-})$ with positive and negative energies
are
\begin{equation}
u(p_-)
=
 {1 \over \sqrt{2\omega_p}}\pmatrix{
{1 \over (1+s)^{1 \over 4}}(\omega_p+p_-)^{1 \over 2} \cr
{i \over (1-s)^{1 \over 4}}(\omega_p-p_-)^{1 \over 2}},
\quad
v(p_-)
=
{1 \over \sqrt{2\omega_p}}\pmatrix{
{1 \over (1+s)^{1 \over 4}}(\omega_p+p_-)^{1 \over 2} \cr
{-i \over (1-s)^{1 \over 4}}(\omega_p-p_-)^{1 \over 2}}.
\label{freespinor}
\end{equation}
These spinors satisfy equations of motion
and 
completeness relations
\begin{equation}
(\gamma^\mu p_\mu-m) u(p) = 0, \qquad
(\gamma^\mu p_\mu+m) v(p) = 0,
\end{equation}
\begin{equation}
u(p_-)\bar u(p_-)={1 \over 2\omega_p}
\left[\gamma^{\mu}p_{\mu}+m\right], \quad
v(p_-)\bar v(p_-)={1 \over 2\omega_p}
\left[\gamma^{\mu}p_{\mu}-m\right].
\end{equation}
The Bogoliubov transformation is an orthogonal transformation
between annihilation operator of quark with momentum $p_-$
and the creation operator of antiquark with momentum $-p_-$
as given in eq.(\ref{bogoliubovtranfop}).
By defining a new spinors $U$ and $V$, we can rewrite the fermion
field in the first line to the second line of eq.(\ref{bogoliubov})
\begin{eqnarray}
U(p_-) &=& \frac{1}{\sqrt{1 + \Phi^2(p_-)}}
\left[ u(p_-) + \Phi(p_-) v(-p_-) \right],
\nonumber \\
V(-p_-) &=& \frac{1}{\sqrt{1 + \Phi^2(p_-)}}
\left[ v(-p_-) - \Phi(p_-) u(p_-) \right] .
\label{bogoluibovspinor}
\end{eqnarray}
%
%
\newcommand{\AP}[1]{{\it Ann.\ Phys.\ }{\bf #1}}
\newcommand{\CMP}[1]{{\it Commun.\ Math.\ Phys.\ }{\bf #1}}
\newcommand{\IJMP}[1]{{\it Int.\ J. Mod.\ Phys.\ }{\bf #1}}
\newcommand{\JETP}[1]{{\it Sov.\ Phys.\ JETP.\ }{\bf #1}}
\newcommand{\JP}[1]{{\it J.\ Phys.\ }{\bf #1}}
\newcommand{\MPL}[1]{{\it Mod.\ Phys.\ Lett.\ }{\bf #1}}
\newcommand{\NP}[1]{{\it Nucl.\ Phys.\ }{\bf #1}}
\newcommand{\PL}[1]{{\it Phys.\ Lett.\ }{\bf #1}}
\newcommand{\PR}[1]{{\it Phys.\ Rev.\ }{\bf #1}}
\newcommand{\PRL}[1]{{\it Phys.\ Rev.\ Lett.\ }{\bf #1}}
\newcommand{\PTP}[1]{{\it Prog.\ Theor.\ Phys.\ }{\bf #1}}
\newcommand{\PTPS}[1]{{\it Prog.\ Theor.\ Phys.\ Suppl.\ }{\bf #1}}
\newcommand{\RMP}[1]{{\it Rev.\ Mod.\ Phys.\ }{\bf #1}}
%

%
%
\section*{Figure Caption}
\begin{itemize}
\item[Fig. \ 1]
Dispersion relations of a free massive particle on (a) usual,
  (b) an interpolating and (c) light-cone quantization surfaces.
\end{itemize}
\end{document}